\documentclass[twocolumn,prl,aps,floatfix,superscriptaddress,amssymb]{revtex4}
\usepackage{epsfig}

\begin{document}

\title{Localization in an Inhomogeneous Quantum Wire}

\author{A.~D.~G\"u\c{c}l\"u}
\affiliation{Department of Physics, Duke University, Box 90305,
Durham, North Carolina 27708-0305}

\author{C.~J.~Umrigar}
\affiliation{Laboratory of Atomic and Solid State Physics, Cornell University, Ithaca, New York 14853}

\author{Hong~Jiang}
\affiliation{Fritz-Haber-Institut der Max-Planck-Gesellschaft, Faradayweg 4-6, D-14195 Berlin, Germany}

\author{Harold~U.~Baranger}
\affiliation{Department of Physics, Duke University, Box 90305,
Durham, North Carolina 27708-0305}

\date{August 5, 2009}

\begin{abstract}
We study interaction-induced localization of electrons in an inhomogeneous quasi-one-dimensional system---a wire with two regions, one at low density and the other high. Quantum Monte Carlo techniques are used to treat the strong Coulomb interactions in the low density region, where localization of electrons occurs. The nature of the transition from high to low density depends on the density gradient---if it is steep, a barrier develops between the two regions, causing Coulomb blockade effects. 
Ferromagnetic spin polarization does not appear for any parameters studied.
The picture emerging here is in good agreement with 
measurements of tunneling between two wires.
\end{abstract}

\maketitle

%
%

With the rapid development of nanotechnology over the last decade, experiments have been able to probe strong interaction phenomena in reduced dimensionality systems such as quantum dots,  wires, and point contacts. Of particular interest are systems in which the electron density is \textit{inhomogeneous}. In the low density region of such a system the interaction energy is comparable to the kinetic energy and novel effects occur such as the ``0.7 structure'' in quantum point contacts or Coulomb blockade effects accompanying localization in a one-dimensional (1D) wire. We perform quantum Monte Carlo calculations of an inhomogeneous, quasi-1D electron system in order to address such effects.

The so-called  ``0.7 structure'' \cite{JPCMissue08} remains poorly understood: in a quasi-1D electron gas -- a wire, constriction, or quantum point contact (QPC) -- decreasing the density causes the conductance $G$ to decrease in integer multiples of 
$2e^2/h$ (one for each transverse mode), except for an extra plateau or shoulder at $G\approx 0.7 (2e^2/h)$ as the lowest mode is depopulated \cite{TNS+96,KBH+00,CLG+02,RBO+02,GTP+03,RPW06,dPP+05}.  Proposed theoretical explanations have been mainly based on three approaches: formation of a bound state leading to a Kondo effect \cite{MHW02,RM06}, spontaneous spin polarization of the low density electrons \cite{TNS+96,RBO+02,Rei05,GTP+03,WB96,RPW06}, and formation of a Wigner crystal \cite{Mat04,Mat04b}. An open question is whether the critical features underlying each of these approaches is present in an inhomogeneous quasi-1D system---whether a localized state with Kondo-like correlations, spin polarization, or a Wigner crystal occurs.

The formation of a Wigner crystal was investigated directly using tunneling spectroscopy into a quantum wire \cite{SAY+06,AYdP+02,ASY+05,FQT+05}. Clear evidence of localized electrons was found, accompanied by unexpected single electron phenomena. 

The general problem of a transition from a liquid to a localized crystal-like phase remains a subject of fundamental research in a variety of bulk and nanoscale systems \cite{KravSarachik04,JameiKivSpiv05,GGU+06,GGU+08,Casula06,Casula08,DB08}. Thus, an approach from a quasi-1D point of view is valuable not only for understanding the 0.7 anomaly and the tunneling experiments, but also for bringing a new way of looking  at the physics of interaction-induced liquid to crystal transitions.

Previous electronic structure calculations investigating inhomogeneous 1D systems have been based on mean field approximations. While some density  functional calculations in the local spin density approximation (LSDA) were used to support the spontaneous spin polarization scenario \cite{WB96,HPN+04}, other LSDA and Hartree-Fock (HF) calculations \cite{MHW02,RM06,Sus03,Mueller05,IZ07} confirmed the existence of quasi-bound states, which may lead to Kondo-like physics in a short QPC \cite{CLG+02,MHW02}. A very recent HF calculation in a long constriction in a weak magnetic field predicted an antiferromagnetic to ferromagnetic transition \cite{QH08}. Despite considerable interest, no many-body calculation beyond mean-field has, as far as we know, been performed for an inhomogeneous 1D system.

Here we use variational and diffusion quantum Monte Carlo (QMC) techniques to investigate correlation and localization in a zero temperature, inhomogeneous, quasi-1D electron system. We observe interaction-induced localization in the low density region.
Our focus is on the transition between the high and low density regions, which can be either smooth or abrupt and may involve the formation of a barrier, depending on the smoothness of the external potential.

%

To study interacting electrons in an inhomogeneous quasi-1D system, we consider a narrow two-dimensional quantum ring with a constriction (point contact):
\begin{eqnarray}
\nonumber
H&=&-\frac{1}{2}\sum_i^N\bigtriangledown_i^2
 +\frac{1}{2}\sum_i^N\omega^2 (r_i-r_0)^2
 +\sum_{i < j}^N {1 \over r_{ij}} \\
 &+& V_g \left\{\mbox{tanh}\left[s(\theta_i+\theta_0)\right]
            -\mbox{tanh}\left[s(\theta_i-\theta_0)\right]\right\}
\label{hamiltonian}
\end{eqnarray}
where effective atomic units are used---the length scale is $a_0^* = \hbar^2 \epsilon/m^* e^2$ and the energy scale is the effective Hartree $H^* = e^2/\epsilon a_0^*$. For comparison with the tunneling experiments, the values for GaAs are $a_0^* = 9.8$\,nm and $H^* = 11.9$\,meV. The parabolicity $\omega$ controls the width of the ring, $r_0$ is the radius, and $V_g$ is the gate voltage that controls the electronic density in the low density constriction. The sharpness and length of the gate potential are tuned through the parameters $s$ and $\theta_0$. In this work, $\omega$ and $r_0$ are set to $0.6$ and $25$; $\theta_0$ is $1.5$ radians. Our potential yields a low density region of length $75$ ($0.73$ $\mu\mbox{m}$ for GaAs). $N=30$ or $31$, generating a single-mode 1D electron density of $n \sim\!20$ $\mu\mbox{m}^{-1}$ when $V_g = 0$. The electron gas parameter, $r_s \equiv 1/2na_0^*$, is $2.6$ in this case.
For previous work on (uniform) quantum rings, 
see Ref.\,\onlinecite{ReimannMannRMP02}.

\begin{figure}
\epsfig{file=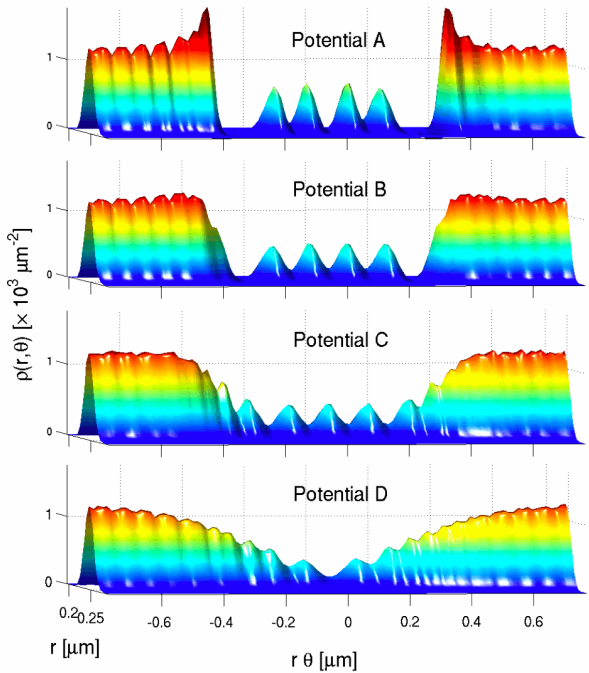,width=3.2in}
\caption{\label{fig:sharp2smooth}
(Color online) Two-dimensional ground state density $\rho$ for gate potentials of different shape [all with $V_g=0.8$ (in units of $H^*$) and $N=31$]. Potentials A, B, and C have $s=15$, $4$, and $2$, re\-spectively. D is obtained using a gaussian with an angular width of 1.2 radians. Potential B is used in most of the paper.
}
\end{figure}

Variational Monte Carlo (VMC) is the first step of our numerical approach. We use recently developed energy optimization methods \cite{UmrigarFilippi05,Umrigar07} to optimize a Jastrow-Slater type trial wave function  $\Psi_T({\bf R})=J({\bf R})D({\bf R})$ \cite{GJU+05}. To build the Slater determinant $D({\bf R})$, we considered three qualitatively different types of orbitals: 
$\chi (r)\, e^{\pm in\theta}$ with $\chi (r)$ fixed,
LSDA orbitals, and floating  gaussians \cite{GGU+08}. Most of the results presented here were obtained using floating gaussians as they provide a better description in the strongly localized regime -- a comparison is given below.
After optimizing the variational parameters (Jastrow parameters as well as, separately, the positions and radial/angular widths of the floating gaussians), we then perform a diffusion Monte Carlo (DMC) calculation to project the trial wave function onto the fixed-node approximation of the true ground state \cite{FMN+01}. The fixed-node DMC energy is an upper bound on the true  energy and depends only on the nodes of $\Psi_T$ obtained from VMC.

\begin{figure}
\epsfig{file=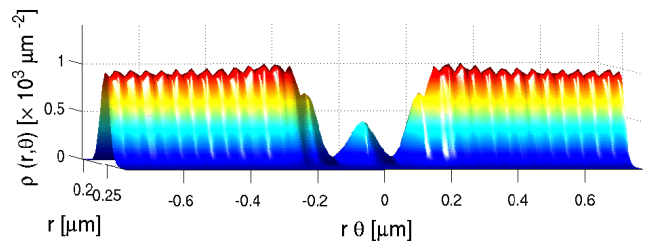,width=3.3in}
\caption{\label{fig:shortconstr}
(Color online) For a short constriction, the two-dimensional ground state density, showing a single localized electron. [$V_g = 0.5$ (in units of $H^*$), $\theta_0 = 0.7$, and $N=30$.]
}
\end{figure}

The density plots of Figs.\,\ref{fig:sharp2smooth} and \ref{fig:shortconstr} give an overview of different scenarios that can occur in an inhomogeneous quantum wire depending on the gate potential landscape. Potential A, which is close to a square barrier, gives rise to three interesting phenomena: (i) The ripples in the high density part of the ring are Friedel oscillations (1 maximum per 2 electrons). This is a signature of weak interactions -- the electrons are in a liquid-like state. (ii) The modulation in the low density part of the wire shows 4 electrons that are individually localized. (iii) There is a large gap separating the liquid and crystal phases, causing the low density region 
to be in the Coulomb blockade regime. This effect has been observed  experimentally, but the origin was not understood \cite{SAY+06,AYdP+02,ASY+05,QH08,FQT+05}. 

For a smoother potential step, the size of the gap decreases (potential B in Fig.\,\ref{fig:sharp2smooth}) and eventually disappears (potential C). Finally, for the gaussian-shaped potential D, localization is very weak. 

As the length of the low density region is reduced, the number of localized electrons decreases. In the extreme case, the low density region becomes like the saddle potential of a quantum point contact. Fig.\,\ref{fig:shortconstr} shows that for a short constriction, a single electron is localized in the low density region, with substantial barriers to the high density ``leads'', as in LSDA calculations \cite{MHW02,RM06,IZ07}. The spin of the localized electron fluctuates, as in the ``Kondo scenario'' \cite{CLG+02,MHW02,RM06} for the 0.7 structure.

The results in Figs.\,\ref{fig:sharp2smooth} and \ref{fig:shortconstr} show that (1) \emph{it is possible to have localized electrons in a low density region distinctly separated from the liquid leads} and (2) \emph{an abrupt potential barrier with a flat plateau enhances both localization and the gap between liquid and crystal regions.}

\begin{figure}
\epsfig{file=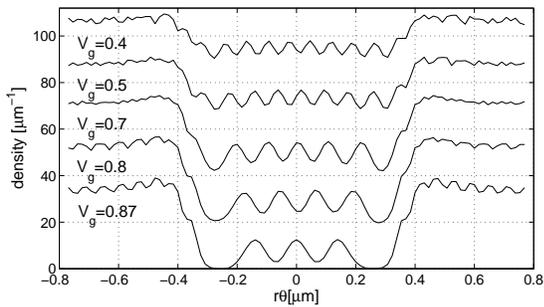,width=3.2in}
\caption{\label{fig:denvsVg}
Electron density as the low density region is depleted. $V_g$ (in units of $H^*$) is varied for potential shape B ($s=4$, $N=31$); each curve is shifted vertically by 20 $\mu$m$^{-1}$ with respect to the one below. For $V_g=0.4$, 8 electrons are localized. As $V_g$ increases, localization becomes stronger and a gap forms at the crystal-liquid boundary---the system is in the Coulomb blockade regime.
}
\end{figure}

Fig.\,\ref{fig:denvsVg} presents how localization develops as the gate potential increases. For the rest of the paper we focus on potential B ($s=4$), for which the width of the potential riser is $\sim 60$ nm. When $V_g=0.4$, the density modulation indicates that localization of electrons is already beginning. The average density at the constriction is $\sim\!15$ $\mu\mbox{m}^{-1}$ ($r_s \sim 3.4$), which is close to but lower than the experimentally estimated critical density $n^*\approx 20$ $\mu\mbox{m}^{-1}$ for localization \cite{SAY+06}. For small $V_g$ ($V_g < 0.2$), no oscillations occur. On the other hand, as $V_g$ increases, the electronic density decreases. The liquid/crystal gap becomes clearly visible at $V_g=0.7$.  At that point 5 individually localized electrons are well separated from the bulk ($r_s \sim 5.7$). Further increases in $V_g$ cause the number of electrons in the constriction to decrease abruptly at certain values while the liquid-crystal gap becomes even stronger. The physical picture that emerges from our QMC calculations is remarkably similar to that in the momentum resolved tunneling experiments \cite{SAY+06,AYdP+02,ASY+05,FQT+05}.

\begin{figure}[b]
\epsfig{file=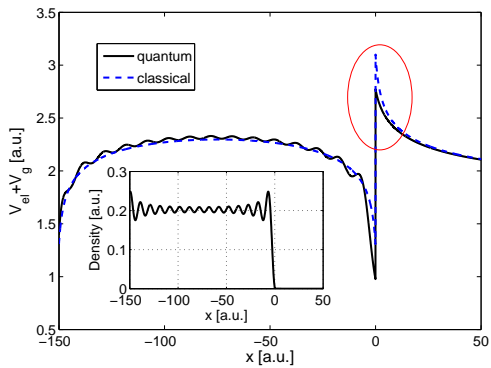,width=0.40\textwidth}
\caption{\label{fig:singlepar}
(Color online) The total effective potential (electrostatic plus external) of bulk electrons in a square well [potential is 1.8 for $x>0$ and $x<-150$, 0 otherwise (other parameters given in \cite{params_singlepart})].  Solid line is from a quantum mechanical single-electron calculation (Inset: density, with Friedel oscillations comparable to those seen in Fig.\,\ref{fig:sharp2smooth} for potential A). Dashed line is obtained assuming a constant density distribution. Clearly, an electron at $x>0$ would feel a potential barrier separating it from the bulk electrons.
}
\end{figure}

To investigate the formation of a barrier between the liquid and crystal regions, consider a model calculation: suppose the gate voltage has depleted the low density region to the point that no electrons reside there. Clearly, an additional electron in the low density region would feel a barrier caused by the electrostatic repulsion of the electrons in the leads. For example, Fig.\,\ref{fig:singlepar} shows for two situations the effective potential due to the repulsion of the bulk electrons plus a square barrier gate potential (finite step at $x=0$). Our QMC results suggest that a barrier is similarly formed even when the density in the constriction is not so depleted, though the discreteness of charge, quantum interference, and correlation probably play a role in enhancing the effect.


\begin{figure}
\epsfig{file=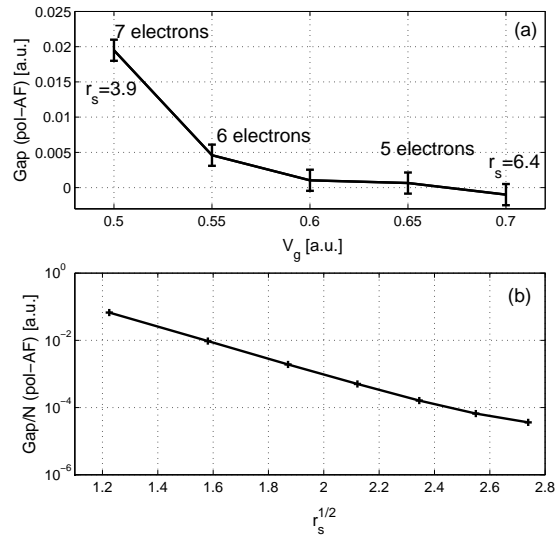,width=0.45\textwidth}
\caption{\label{fig:spingap}
(a) Difference of total energy when electrons in the constriction are fully polarized or antiferromagnetically ordered, as a function of $V_g$ ($N=31$). The antiferromagnet is lower in energy for $V_g<0.6$.
(b) For a homogeneous ring with 30 electrons, energy difference per electron between fully polarized and antiferromagnetic states. The antiferromagnetic state is always favored for the parameters we have studied. ($V_g = 0$; density varied by changing $r_0$.)
}
\end{figure}

The spin structure of the electrons in the low density region is an important physical property which has been controversial \cite{JPCMissue08,TNS+96,RBO+02,Rei05,GTP+03,WB96,Mat04,Mat04b,RPW06,QH08}. In one dimension, the ground state cannot be ferromagnetically polarized (Lieb-Mattis theorem \cite{LM62}), but the transverse degree of freedom may change the situation. With our method, a rigorous treatment of the spin is technically difficult: 
building eigenfunctions of ${\bf S}^2$ out of floating gaussian orbitals for a large number of electrons with $S \ll N/2$ requires a huge number of determinants. We can solve this problem by doubly occupying extended orbitals (i.e., $\chi (r)\, e^{\pm in\theta}$ or LDA orbitals if a self-consistent solution can be found), but this is significantly less accurate than floating gaussians. 
Thus, here we present results using $S_z$ conserving floating gaussian trial wave functions. 

First, we find the energy difference for a homogeneous ring (i.e., $V_g = 0$) between fully polarized and antiferromagnetic states. Fig.\,\ref{fig:spingap}(b) shows that the ferromagnetic state is always higher in energy and that the difference decays exponentially in $\sqrt{r_s}$, as expected \cite{Mat04,Mat04b}.

In order to compare to experiment \cite{SAY+06}, we also study the spin state in the inhomogeneous ring. Fig.\,\ref{fig:spingap}(a) shows the energy difference between ferromagnetically and antiferromagnetically arranged electron spins in the constriction. (In the high density ``leads'', the spins have no definite arrangement because of strong fluctuations coming from weak correlations.) At $V_g=0.5$, the low density part is clearly \textit{not} fully spin polarized as the spin gap is as large as $\sim 3$ K. However, as $V_g$ increases, both the number of  electrons and their overlap decrease rapidly, resulting in a much smaller energy gap. For $V_g>0.6$, the energy gap is not resolved due to statistical error of about $0.5$ K. This is consistent with Steinberg {\it et al.}'s experiments performed at temperatures down to $\sim\! 0.3$ K \cite{SAY+06}: it was found that for $N<6$ the observed state is a mixture of ground and thermally excited spin states.

\begin{figure}
\epsfig{file=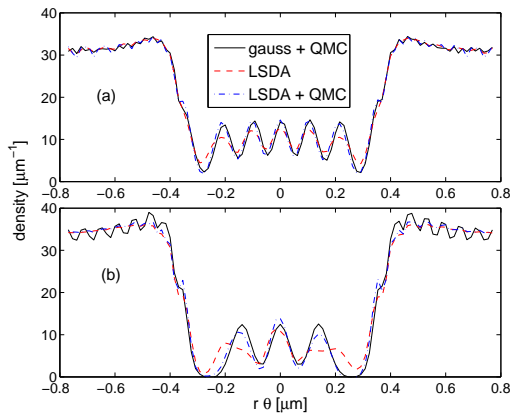,width=0.4\textwidth}
\caption{\label{fig:ldavsqmc}
(Color online) Comparison of ground state densities obtained using LSDA, QMC with
LSDA orbitals, and QMC with gaussian orbitals at (a) $V_g=0.7$ and (b) $V_g=0.87$.
}
\end{figure}

We close by comparing the results obtained with different methods and trial wave functions. Fig.\,\ref{fig:ldavsqmc} shows the densities obtained from LSDA and from two QMC calculations, one using LSDA orbitals in the trial wavefunction and the other using floating gaussian orbitals. In Fig.\,\ref{fig:ldavsqmc}(a) there is excellent agreement between the two QMC calculations whereas the LSDA density is not as localized. Fig.\,\ref{fig:ldavsqmc}(b) shows that the inaccuracies of LSDA 
become more visible at higher $V_g$ (lower density), as expected. The gap that forms between the localized and liquid regions is bigger in QMC than in LSDA, indicating a correlation contribu\-tion to the gap. We also performed QMC calculations using $\chi (r)\, e^{\pm in\theta}$ orbitals (hence fixing the total spin), but the density is substantially different and the energy considerably worse. Finally, although the fixed-node DMC energies obtained from LSDA and gaussian orbitals are very close [51.7915(6) and 51.7862(2) respectively at $V_g=0.87$], the floating gaussian based trial wave functions yield significantly reduced fluctuations of the local energy (standard deviation of 0.45 compared to 0.59). Thus floating gaussian trial wave functions provide a better physical description of the system in the Wigner crystallized regime.


We thank O.\ M.\ Auslaender, M.\ Casula, and K.\ A.\ Matveev for helpful discussions. This work was supported in part by the NSF (DMR-0506953, DMR-0205328) and the DOE-CMSN (DE-FG02-07ER46365).

\vspace*{-0.22in}


\end{document}